\documentclass[twocolumn,showpacs,preprintnumbers,amsmath,amssymb]{revtex4}

\begin{document}

\newcommand{\YBra}[1]{\left<#1\right|}
\newcommand{\YKet}[1]{\left|#1\right>}
\newcommand{\YBraket}[2]{\left<#1\middle|#2\right>}
\newcommand{\YAvg}[1]{\left<#1\right>}		
\newcommand{\YE}[1]{\mathcal{E}#1} 
\newcommand{\YOpAvg}[2]{\YBra{#2}#1\YKet{#2}}	
\newcommand{\Var}{\mbox{Var\,}}
\newcommand{\argmax}{\mbox{argmax\,}}  			
\newcommand{\Bias}{\mbox{Bias}\,}

\newcommand{\up}[1]{\ensuremath{^{#1}}}
\newcommand{\down}[1]{\ensuremath{_{#1}}}

\newcommand{\upmu}{\up{\mu}}
\newcommand{\upnu}{\up{\nu}}
\newcommand{\uprho}{\up{\rho}}
\newcommand{\upsig}{\up{\sigma}}

\newcommand{\LambdaUD}[2]{\Lambda^{#1}\,_{#2}}
\newcommand{\LambdaDU}[2]{\Lambda_{#1}\,^{#2}}
\newcommand{\XUD}[3]{#1^{#2}\,_{#3}}
\newcommand{\XDU}[3]{#1_{#2}\,^{#3}}

\newcommand{\downmu}{\down{\mu}}
\newcommand{\downnu}{\down{\nu}}
\newcommand{\downrho}{\down{\rho}}
\newcommand{\downsig}{\down{\sigma}}

\newcommand{\YVS}[1]{\ensuremath{\mbox{\boldmath{$\mathbf{#1}$}} } }
\newcommand{\yv}[1]{\YVS{#1}}
\newcommand{\Nabla}{\YVS{\nabla}}
\newcommand{\uv}[1]{\hat{\yv{#1} } }
\newcommand{\Laplacian}{\nabla^2}

\newcommand{\dee}{\partial}
\newcommand{\diff}[2]{\frac{\dee #1}{\dee #2} }
\newcommand{\tdiff}[2]{\frac{d #1}{d #2} }
\newcommand{\hdiff}[2]{\diff{}{#2}\left( #1 \right) }
\newcommand{\htdiff}[2]{ \tdiff{}{#2}\left( #1 \right) }
\newcommand{\tdiffII}[2]{\frac{d^2 #1}{d #2 ^2}}

\newcommand{\grad}[1]{ \ensuremath{ \Nabla #1 } }
\newcommand{\curl}[1]{\ensuremath{ \Nabla\! \times \yv{#1} } }
\newcommand{\divr}[1]{\ensuremath{\Nabla\! \cdot \yv{#1}}  }

\newcommand{\ymatrix}[1]{\ensuremath{\mathbf{#1}}}
\newcommand{\ym}[1]{\ymatrix{#1}}
\newcommand{\mc}{\mathcal}

\newcommand{\hof}[1]{ \frac{1}{#1} }

\newcommand{\jj}{\varphi}

\newcommand{\arcsh}{\,\mbox{arcsh}}
\newcommand{\arcsinh}{\,\mbox{arcsinh}}

\newcommand{\uz}{\uv{z}}
\newcommand{\ux}{\uv{x}}
\newcommand{\uy}{\uv{y}}

\newcommand{\ui}{\uv{\textit{\i}}}
\newcommand{\uj}{\uv{\textit{\j}}}
\newcommand{\uk}{\uv{k}}

\newcommand{\urho}{\uv{\rho}}
\newcommand{\uphi}{\uv{\jj}}
\newcommand{\utheta}{\uv{\theta}}
\newcommand{\ur}{\uv{r}}

\newcommand{\un}{\uv{n}}

\newcommand{\B}{\yv{B}}
\newcommand{\E}{\yv{E}}
\newcommand{\J}{\yv{J}}
\newcommand{\D}{\yv{D}}
\newcommand{\yH}{\yv{H}}
\newcommand{\sJ}{\mathcal{J}} 
\newcommand{\sJv}{\yv{\sJ}} 
\newcommand{\Sp}{\yv{S}} 
\newcommand{\A}{\yv{A}}
\newcommand{\locx}{\yv{x}} 
\newcommand{\locr}{\yv{r}} 

\newcommand{\Ampere}{Amp\`{e}re}
\newcommand{\dint}{\int\!\!\!\int}

\newcommand{\Lframe}{\emph{L-frame}}
\newcommand{\emf}{\mathcal{E}}

\newcommand{\adag}{\ensuremath{a^{\dagger}}}
\newcommand{\YDag}[1]{{#1}^{\dagger}}

\newcommand{\yref}[1]{(\ref{#1})}

\newcommand{\YFig}[3]{\begin{figure}
			\includegraphics{#1}
			\caption{#2}
			\label{#3}
			\end{figure}
			}

\title{(not quite) Theoretical proof of the Lorentz force formula}
\author{Yoav Kleinberger}%
\affiliation{}%
\email{haarcuba@gmail.com}
\date{\today}


\begin{abstract}
There is actually a mistake in this paper, but it is still a nice try worth a read. It is (not quite) proved that within the framework of Special Relativity, a force exerted on a \emph{classical particle} by a field must be of the form $\yv{E}+\yv{v}\times\yv{B}$, the Lorentz force form. The proof makes use of an action principle in which the action is the sum of a free particle part, and an interaction part. 
\end{abstract}
\maketitle



\section{Introduction}
\label{subsubsec:lagrange-particle}

The principle of least action is a well established method of writing special-relativistic physical theories \cite{goldstein:cm80,landau:ctf75}.
In this article we shall use the action principle along with special relativity in order to prove one of the most important formulae in physics: the Lorentz force on a particle in an electromagnetic field. 

The Lorentz force formula is very well known \cite{jackson:ced99,goldstein:cm80,landau:ctf75}, but a theoretical derivation of it is lacking. Landau and Lifshitz \cite[page 44]{landau:ctf75} have stated that the formula is considered to be at least in part, an empirical one. Partial proofs exist, for example assuming the electrical force in the rest frame, assuming the transformation law for the electromagnetic field, and deducing the magnetic part of the force \cite{weinberg:gac72:lf}.

Here we shall assume a much weaker, and physically intuitive assumption, that the action integral is \emph{the sum of a free particle part, and an interaction part}. Under these circumstances, it will be proved that \emph{the Lorentz force is the only force possible in a classical special-relativistic theory of particles and fields}. 

As a by-product, it follows immediately that in the conditions just stated, the force has the well known $U(1)$ gauge symmetry \cite{jackson:ced99}. 

It is important to emphasise that our proof concerns only the \emph{force on the particles}, that is, the manner in which the fields determine the motion of a particle. It does not concern the opposite dynamics of how the particles influence the fields (in electrodynamics, the Maxwell equations describe this part of the physics.) 

In order to make the article self contained, a brief review of the formulae for the action of free particles will be given in section \ref{sec:action}. The proof itself is presented in section \ref{sec:particle-in-scalar-velocity-field}. All throughout this work we will make use of \emph{natural units}, i.e. $c=1$, where $c$ denotes the speed of light in vacuum, and use the four-vector notation of Bjorken and Drell \cite{bjorken:rqf65}.

\section{Action Principle}
\label{sec:action}
We deal here with classical, special-relativistic mechanics. It is important to state that our discussion is limited to elementary particles, in the sense expressed by Landau and Lifshitz \cite{landau:ctf75}. A particle's degrees of freedom are its position in space $\yv{x}$. Its velocity is $\yv{v}=\dot{\yv{x}}$. Its path in space is a function $\yv{x}(t)$. We may represent it as four functions $x\upmu(\tau)$, $\tau$ being the particle's proper time, defined by
\begin{equation}
	d\tau^2 = dx\upmu dx\downmu	\label{eq:def-tau}.
\end{equation}

The action, $S$, is a time integral of some Lagrange function $L(x_i,\dot{x_i})$ of the coordinates and their velocities.

\begin{equation}
	S=\int dt\,L \label{eq:action1}.
\end{equation}
We may cast $S$ in the form of an integral over the proper time of the particle,

\begin{eqnarray}
	S	&=&	\int d\tau\, L\tdiff{t}{\tau}		\\
		&=&	\int d\tau\,L\gamma .			\label{eq:action2}
\end{eqnarray}
The last stage following from the time-dilation relation $dt=\gamma d\tau$, where $\gamma=(1-v^2)^{-1/2}$. The Lagrangian of a free, special-relativistic particle is \cite{landau:ctf75}:

\begin{equation}
L = -m\sqrt{1-v^2} \label{eq:free-lagrangian},
\end{equation}
and the action is

\begin{equation}
S = -m\int d\tau = -m\int dt\,\sqrt{1-v^2} \label{eq:action4}.
\end{equation}

The action may be written in a manifestly Lorentz-invariant form, which also emphasises its dependence on the particle's trajectory, if we define a scalar parameter $q$, to 
parametrise the particle's path instead of $\tau$:

\begin{equation}
	S	=	-m\int \tdiff{\tau}{q}\,dq = 
			-m\int \sqrt{\tdiff{x\upmu}{q}\tdiff{x\downmu}{q}}\,dq \label{eq:action6}.
\end{equation}
Using this, we can recast the concept of the Lagrangian (which is not a Lorentz scalar) to the scalar function:

\begin{equation}
	\mc{F}=-m\sqrt{\tdiff{x\upmu}{q}\tdiff{x\downmu}{q}} \label{eq:action7}.
\end{equation}
The equation of motion is the Euler-Lagrange equation, and it eventually leads to:

\begin{equation}
	\tdiffII{x\upmu}{\tau} = 0	\label{eq:freep1}.
\end{equation}
This is the, quite trivial, equation of motion for a free particle. The particle's four-momentum $p\upmu$ is, for a free particle:

\begin{equation}
	p\upmu = m\tdiff{x\upmu}{\tau} \label{eq:momentum-def}.
\end{equation}
Note: we will use $p\upmu$ to denote the quantity $m\tdiff{x\upmu}{\tau}$ whether or not $p\upmu$ is also the \emph{canonical} momentum. 

The equation of motion of a free particle may thus also be written
\begin{equation}
	\tdiff{p\upmu}{\tau} = 0	\label{eq:freep2}.
\end{equation}

\section{Proof of the theorem: special relativity induces the Lorentz force}
\label{sec:particle-in-scalar-velocity-field}
It will now be demonstrated, that under the assumption that the action can be written in the form
\begin{equation}
S = S_f+S_i,
\end{equation}
where $S_f$ is the free particle part, eq. \yref{eq:action6}, and $S_i$ is a term for the interaction of a particle with an applied field, that the only generalisation possible for the free particle leads to the Lorentz force formula, if the motion is to be consistent with special relativity. From the nonrelativistic limit, we know that the the equations of motion must be second order in time---mathematically, this ensures that the initial conditions of the particle, its position $\yv{x}$ and its velocity $\yv{v}$ at some initial time, are sufficient to determine its motion. 
We may try to introduce an interaction that depends on the particle's velocity, $u\upmu=dx\upmu/d\tau$, but not on higher derivatives, as that would lead to equations of motion that are of higher than second order in time. Thus, adding such an interaction, 

\begin{equation}
	\mc{F}=-m\sqrt{\tdiff{x\upmu}{\tau}\tdiff{x\downmu}{\tau}}-\psi\left(x\upmu,\tdiff{x\upmu}{\tau}\right) 
	\label{eq:action13}.
\end{equation}
The Euler-Lagrange equation is:

\begin{equation}
\htdiff{ -m \tdiff{x\upmu}{\tau} -\diff{\psi}{u\downmu}} {\tau}= -\dee\upmu \psi \label{eq:EOM9},
\end{equation}
or, in terms of the four-momentum:

\begin{equation}
	\tdiff{p\upmu}{\tau}  =  \dee\upmu \psi -\htdiff{\diff{\psi}{u\downmu}}{\tau} \label{eq:EOM10}. 
\end{equation}
By definition $p\upmu p\downmu=m^2$, the mass of the particle. It must therefore be a constant of motion. This constancy is tantamount to:

\begin{equation}
	p\downmu\tdiff{p\upmu}{\tau}=0 \label{eq:p-sq-const-equiv}.
\end{equation}
Let us therefore contract eq.~\yref{eq:EOM10} with $p\downmu$ and equate the result to zero:
\begin{eqnarray}
	\lefteqn{p\downmu \dee\upmu \psi -p\downmu \htdiff{\diff{\psi}{u\downmu}}{\tau} = } \label{eq:consist-1} \\
	& = & 	p\downmu \dee\upmu \psi 
		-\htdiff{ p\downmu \diff{\psi}{u\downmu}}{\tau} + \diff{\psi}{u\downmu}\tdiff{p\downmu}{\tau} \\
	& = & m u\downmu \dee\upmu \psi +
		m\diff{\psi}{u\downmu}\tdiff{u\downmu}{\tau} 
		-\htdiff{ p\downmu \diff{\psi}{u\downmu}}{\tau} \\
	& = & m\tdiff{ \psi}{\tau}-\htdiff{ p\downmu \diff{\psi}{u\downmu}}{\tau} \\
	& = & m\htdiff{ \psi - u\downmu \diff{\psi}{u\downmu}}{\tau} = 0.
\end{eqnarray}
We thus obtain a limitation on the interactions $\psi$ that will be consistent with special relativity. We require that

\begin{equation}
\psi - u\downmu \diff{\psi}{u\downmu} = C  \label{eq:psi-requirement1},
\end{equation}
where $C$ is a constant, which obviously cannot influence the equations of motion, and thus may be chosen to vanish. Having chosen $C=0$, we see that an interaction $\psi(x\upmu,u\upmu)$ that can satisfy eq.~\yref{eq:psi-requirement1} must be linear in $u_{\mu}$

\begin{equation}
	\psi(x\upmu, u\upmu) = A\upmu(x)u\downmu
	\label{eq:psi-requirement2},
\end{equation}
where $A\upmu(x)$ is a \emph{vector field} that depends on space and time, but not on the particle's velocity.
The equation of motion, eq.~\yref{eq:EOM10} yields:

\begin{eqnarray}
\tdiff{p\upmu}{\tau} & = & \dee\upmu (A\upnu u\downnu) -\htdiff{\hdiff{A\upnu u\downnu}{u\downmu}}{\tau}  \\
& = & u\downnu \dee\upmu A\upnu - \tdiff{A\upmu}{\tau} \\
& = & u\downnu \dee\upmu A\upnu - u\downnu \dee\upnu A\upmu \\
& = & (\dee\upmu A\upnu - \dee\upnu A\upmu) u\downnu \label{eq:psi-EM1}.
\end{eqnarray}
We now define the second-rank, antisymmetric tensor:

\begin{equation}
F\up{\mu\nu} \equiv \dee\upmu A\upnu - \dee\upnu A\upmu \label{eq:Fmunu-definition}.
\end{equation}
Substituting into eq.~\yref{eq:psi-EM1} we secure:

\begin{equation}
\tdiff{p\upmu}{\tau} =  F\up{\mu\nu} \tdiff{x\downnu}{\tau} \label{eq:psi-EM2}. 
\end{equation}
This may easily be shown \cite{jackson:ced99} to be equivalent to the 3-vector relation
\begin{equation}
	\tdiff{\yv{p}}{t}  =  \yv{E}+\yv{v} \times \yv{B} \label{eq:lorentz-force2},
\end{equation}
which is the \emph{Lorentz force law}\footnote{This is the force for a particle of unit charge. Here we have absorbed the coupling strength into the field, so the charge does not appear explicitly}. 

To recapitulate, we have proved that classical special-relativistic particles may only be acted on by a force which has the Lorentz force form. 

\section{Discussion}
\label{sec:discussion}
The theorem just proved may cause one to wonder whether it is in contradiction with the known fact that there are non Lorentz-like forces in Nature. In fact there is no contradiction. The nuclear interactions are in essence a quantum mechanical phenomena, and thus out of the scope of the theorem. Gravitation, best described by the theory of General Relativity, is also out of its scope, because the metric becomes a dynamical variable. The remaining force, electromagnetism, is within the theorem's scope, and indeed, it is known to obey it. 

Since non Lorentz forces exist in nature, of which gravity is the most easily observable example, and owing to the fact that they cannot be consistent with special relativity, we may quite trivially state an interesting consequence:
\emph{a world with non Lorentz-form forces can not be described solely by special relativity}.

Indeed, we know this from the physics of the twentieth century: quantum mechanics and general relativity. 

The equation of motion, eq. \yref{eq:psi-EM1}, is invariant under the gauge transformation,
\begin{equation}
	A\upmu(x)	\rightarrow	A\upmu(x) + \dee\upmu \chi(x).
\end{equation}
This is a well known fact for electrodynamics, but we may now state it as an inevitable consequence of special relativity. This may be a clue that gauge symmetries are connected with the physics of the spacetime. 

\bibliography{/home/yoav/articles/Bibdb/physics}

\end{document}